\documentclass[twocolumn,aps,showpacs,pre,amsmath]{revtex4-2}

\usepackage{amssymb}
\usepackage{graphicx}
\usepackage{dsfont}
\usepackage{color,soul}

\begin{document}
	
	\title{Electronic localization in small-angle twisted bilayer graphene}
	
	\author{V. Hung Nguyen$^1$,} 
	\author{D. Paszko$^1$, M. Lamparski$^2$, B. \surname{Van Troeye}$^2$, V. Meunier$^2$, and J.-C. Charlier$^1$}
	\affiliation{$^1$Institute of Condensed Matter and Nanosciences, Universit\'{e} catholique de Louvain (UCLouvain), B-1348 Louvain-la-Neuve, Belgium \\
		$^2$Department of Physics, Applied Physics, and Astronomy, Rensselaer Polytechnic Institute, Troy, New York 12180, United States}
	
	\begin{abstract}
	Close to a magical angle, twisted bilayer graphene (TBLG) systems exhibit isolated flat electronic bands and, accordingly, strong electron localization. TBLGs have hence been ideal platforms to explore superconductivity, correlated insulating states, magnetism, and quantized anomalous Hall states in reduced dimension. Below a threshold twist angle ($\sim$ $1.1^\circ$), the TBLG superlattice undergoes lattice reconstruction, leading to a periodic moir\'e structure which presents a marked atomic corrugation. Using a tight-binding framework, this research demonstrates that superlattice reconstruction affects significantly the electronic structure of small-angle TBLGs. The first magic angle at $\sim$ $1.1^\circ$ is found to be a critical case presenting globally maximized electron localization, thus separating reconstructed TBLGs into two classes with clearly distinct electronic properties. While low-energy Dirac fermions are still preserved at large twist angles $> 1.1 ^\circ$, small-angle ($\lesssim 1.1^\circ$) TBLG systems present common features such as large spatial variation and strong electron localization observed in unfavorable AA stacking regions. However, for small twist angles  below $1.1 ^\circ$, the relative contribution of the local AA regions is progressively reduced, thus precluding the emergence of further magic angles, in very good agreement with existing experimental evidence.
	\end{abstract}
	\maketitle
	
	Ever since its first successful controlled isolation \cite{novo04}, graphene has become the subject of intense research, that has since been extended to other similar materials \cite{ferr15}. For example, twisted bilayer graphene (TBLG) is obtained when two graphene monolayers are rotated relative to one  another by an arbitrary angle $\theta$ \cite{andr20,wang19}. TBLGs display many fascinating properties that can be tuned by varying the twist angle $\theta$ \cite{wang19,carr20}. This possibility has given rise to the nascent field of \textit{twistronics} \cite{macd19}. In particular, in the vicinity of the first magic angle experimentally observed at $\sim$ $1.1^\circ$ \cite{bist11,cao18a,yank19,utam21}, strong interlayer hybridization leads to the emergence of localized flat bands near the Fermi level. Because the kinetic energy contribution of these states is negligible, they are prone to give rise to strongly enhanced many-body effects \cite{choi18,kere19}. This strong electronic correlation in small-angle TBLG has led to the observation of several novel phenomena such as superconductivity, correlated insulating states, magnetism, and even quantized anomalous Hall states \cite{cao18a,yank19,kere19,cao18b,lu2019,jian19,xie019,choi19,leon20,serl20,chma20}.

	In spite of the large body of research devoted to the electronic properties of TBLGs as a function of twist angle \cite{sant07,tram10,moon12,hung15,leco19,carr19,grig19,yren21,carr20,wale20,cant20}, a full understanding of the physics governing the emergence of magic angles is still incomplete. This difficulty is due in large part to the presence of a large number of atoms in the supercell of small-angle TBLGs, which precludes a detailed theoretical modeling of their properties\cite{lamp20}. For example, a number of studies using a continuum Dirac Hamiltonian \cite{bist11,grig19,yren21},  predict the existence of a discrete series of other magic angles $< 1.1^\circ$, in addition to the first magic angle located around $1.1^\circ$.  However, these \textit{smaller magic angles} are not predicted by more accurate calculations based on atomistic tight-binding models parameterized by density functional theory \cite{carr19,leco19,wale20}. In fact, the presence of the first magic angle has been demonstrated in several experiments \cite{cao18a,yank19,kere19,utam21} but there is, to-date, no experimental evidence pointing to the existence of other magic angles $< 1.1^\circ$.
	
	Small-angle TBLG systems have been found to be characterized by a significant atomic reconstruction \cite{garg17,nnam17,hyoo19}. In that sense, TBLGs can be considered as a network of topological domain walls (\textit{i.e.}, soliton lines) connecting unfavorable AA stacking regions and separating more energetically favorable triangular AB/BA stacking domains alternatively arranged \cite{efim18,lamp20}. Thus, these small-angle TBLG lattices exhibit a very different stacking network, compared to large-angle ones where the moir\'e superlattice evolves smoothly. As a consequence, a distinguished feature is the observation of low-energy helical edge states in small-angle TBLGs \cite{jose13,huan18,rick18,sgxu19,beul20}. Moreover, this specific stacking structure has been shown to induce a strongly spatial dependence of the electronic and vibrational properties as well as related features \cite{sunk18,lamp20,gade20}. On this basis, an accurate analysis of both the global and local properties of small-angle TBLG systems is urgently needed towards an improved understanding of their unusual electronic properties.
	
	In this research, a tight-binding approach is employed to compute the electronic properties of TBLGs around and below the first magic angle ($\theta^{\rm MA}_1 \simeq 1.1^\circ$). In contrast to most of the previous reports, this work fully takes the moir\'e superlattice atomic reconstruction into account. Interestingly, $\theta^{\rm MA}_1$ is found to be a critical case where the electron localization is maximum, thus separating TBLG systems into two main classes (\textit{i.e.}, large- and small-angles) with clearly distinct electronic features. Remarkably, for angles below $\theta^{\rm MA}_1$, the local electronic properties in AA stacking regions commonly display a strong electron localization, similar to that globally observed for $\theta^{\rm MA}_1$, and are found to reach a saturation when further reducing $\theta$. Such an electron localization does not occur in the AB and BA regions. Instead, the electronic properties of AB/BA stacking zones approach those of Bernal stacking bilayer graphene and helical edge states are observed at low energies around the soliton lines (SL). On the basis of these properties and, including the fact that the contribution of AA zones to the global properties of the system is reduced when $\theta$ decreases, isolated flat bands as observed at $\theta^{\rm MA}_1$ can no longer be obtained for smaller angles. In agreement with experiment (\textit{e.g.},  Refs.~\cite{cao18a,huan18,kere19,gade20,brih12,hude18,zhan20}), our numerical analysis demonstrates that the global electron localization is monotonically reduced when reducing $\theta$ below $\theta^{\rm MA}_1$, thus confirming the absence of smaller magic angles.
	
	TBLG moir\'e superlattices were relaxed to minimize their energy using a second-generation REBO potential \cite{bren02} for intralayer interactions and the registry-dependent Kolmogorov-Crespi potential \cite{kolm05} to describe the interlayer ones \cite{lamp20}. This way, all atomic positions and superlattice parameters of the considered TBLG structures were optimized until all force components are less than 1~meV/atom. 
		Calculations with the mentioned potentials model quite accurately the structural properties of TBLG, i.e., in good agreement with DFT predictions \cite{uchi14,cant20} (see in section 2 of Supplementary Information (SI)) as well as experimental observations (discussed in more detail below).
	The electronic properties of these relaxed lattices were then computed using a standard $p_z$ tight-binding model \cite{tram12}. Note that the tight-binding parameters used in the present work have been adjusted (see the SI) in order to model more accurately the first magic angle experimentally observed at $\sim$ $1.1^\circ$ \cite{andr20,cao18a}. To overcome the numerical challenges related to the very large number (\textit{i.e.}, $10^4$ to $> 10^5$) of atoms in the large supercell of small-angle TBLG, the tight-binding Hamiltonian was solved using recursive Green's function techniques \cite{thor14} that enable computing the real-space and momentum-dependent local density of electronic states (see also the SI). This technique presents the advantage of allowing the determination of both global and local electronic properties.
	
	\begin{figure}[!t]
	\centering
	\includegraphics[width = 0.49\textwidth]{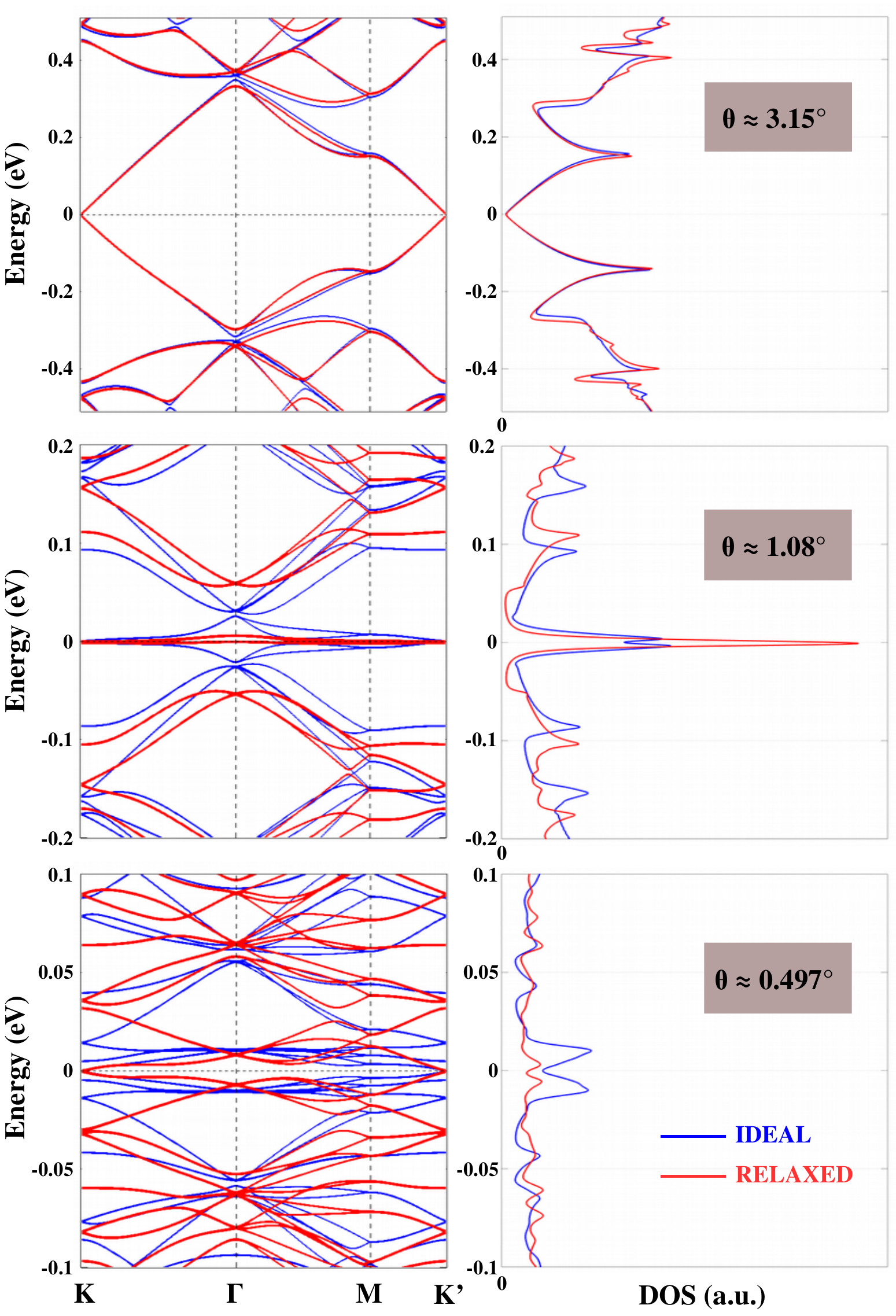}
	\caption{Electronic band structures and corresponding DOS of TBLG systems with different twist angles: above ($\theta \simeq 3.15^\circ$), in the vicinity of ($\theta \simeq 1.08^\circ$) and below ($\theta \simeq 0.50^\circ$) the first magic angle, with (red curves) and without (blue curves) atomic structure relaxation of the moir\'e superlattice.}
	\label{fig_sim1}
	\end{figure}

	\begin{figure*}[!t]
	\centering
	\includegraphics[width = 0.98\textwidth]{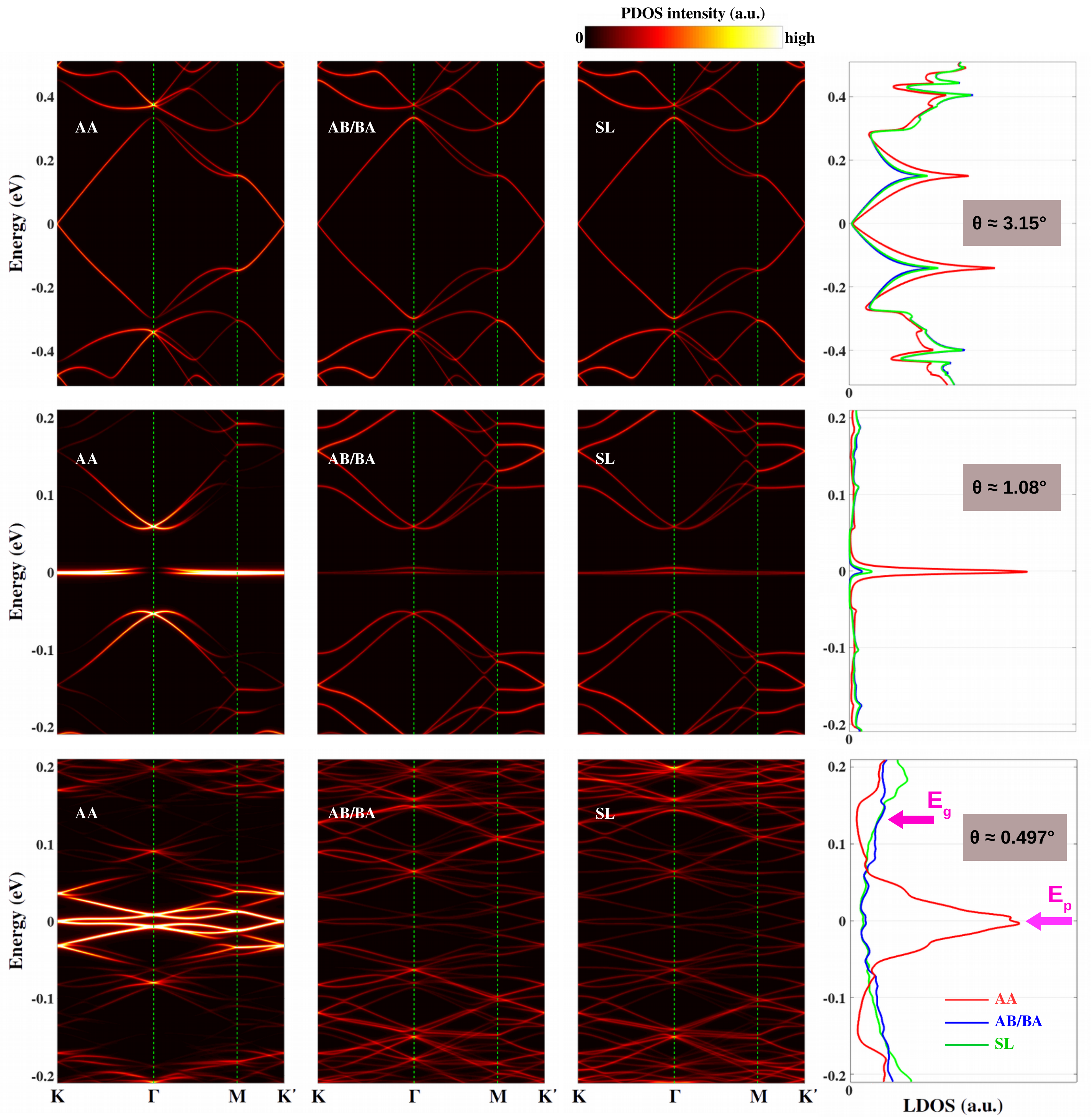}
	\caption{Projected DOS computed in the AA domains (first column), AB/BA domains (second column) and at the center of soliton lines (third column); along with the corresponding local DOS in TBLG for different twist angles as investigated in Fig.~1.}
	\label{fig_sim2}
	\end{figure*}
	Our results confirm that atomic reconstruction is negligible for large twist angles but is essential for small ones since it significantly modifies the stacking structure of TBLGs, compared to the ideal isolated graphene lattices. As a consequence, the electronic properties of small-angle TBLG systems are expected to be strongly affected as well \cite{nnam17,leco19,cant20}. To confirm this statement, electronic band structures and DOS of TBLG featuring three different twist angles ($\simeq 3.15^\circ, 1.08^\circ,$ and $0.50^\circ$) are presented in Fig.~1. For twist angles above $\theta^{\rm MA}_1$, the effect of atomic reconstruction is almost negligible whereas around and below $\theta^{\rm MA}_1$, the electronic properties drastically change when the superlattice relaxation is performed. 
		The atomic corrugation has been studied in Ref.~\onlinecite{luci19} where its role on the electronic properties has been investigated, particularly, regarding the emergence of the flat bands and energy gaps between them and high-energy ones of TBLG at $\sim 1.1^\circ$. In general, as presented and discussed in Sections 2 and 3 of the SI, not only out-of-plane displacements but also in-plane ones are found to be significant and therefore have strong effects on the electronic properties in the small ($\lesssim 1.1^\circ$) angle regime.
	Around $\theta^{\rm MA}_1$ (\textit{i.e.}, $1.08^\circ$ here), the electronic structure of reconstructed TBLG clearly exhibits isolated flat bands and strong electron localization peaks in the DOS near the Fermi level, in very good agreement with experiments and with previous theoretical works \cite{utam21,cao18a,nnam17,leco19,cant20}. For angles below $\theta^{\rm MA}_1$, while the electronic structure of ideal lattices still presents some partially flat bands and corresponding DOS peaks around the Fermi level, all low-energy bands of relaxed TBLG systems are found to be much more dispersive. Note that in all the reconstructed TBLG systems examined in this work (as more fully presented in Section 4 of the SI for $\theta$ down to $\sim 0.22^\circ$), neither isolated flat bands nor highly localized DOS peaks around the Fermi level observed for $\theta^{\rm MA}_1$ occur again for small angles $\theta < \theta^{\rm MA}_1$. 
		This feature can be rationalized by observing that the total DOS (shown in Figs.~1 and 4) exhibits a gradual reduction of electron localization when decreasing the twist angle below $1.1^\circ$. This observation is more fully explored in Figs.~S6a and S6b of the SI.

	\begin{figure}[!t]
	\centering
	\includegraphics[width = 0.49\textwidth]{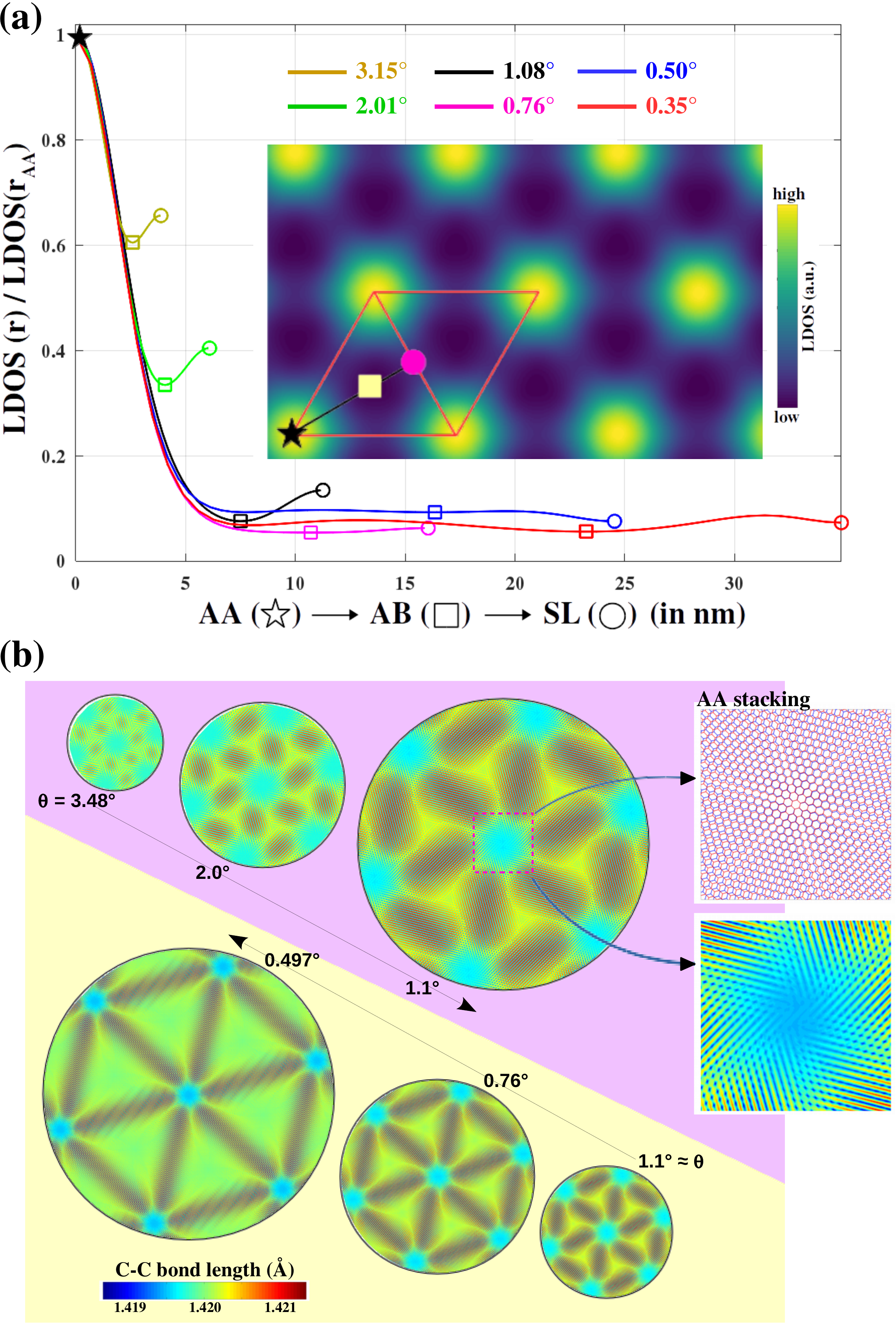}
	\caption{(a) Local DOS extracted along the line from a AA zone (star) to the center of soliton line (circle) passing though AB/BA domain (square). The inset illustrates a real-space image of zero-energy local DOS at $\theta \approx 0.76^\circ$. The local DOS is normalized to its maximum value obtained in the AA domain. Energy at van Hove singularity points of the DOS is considered for angles $> 1.1^\circ$ while zero energy is computed for angles $\lesssim 1.1^\circ$. (b) Spatial variation of the C-C bond length computed for one layer of TBLG with large (top) and small (bottom) twist angles. Insets on top illustrate the strong correlation between weakly strained (light blue) and AA stacking domains.}
	\label{fig_sim3}
	\end{figure}	
	
	\begin{figure}[!t]
	\centering
	\includegraphics[width = 0.49\textwidth]{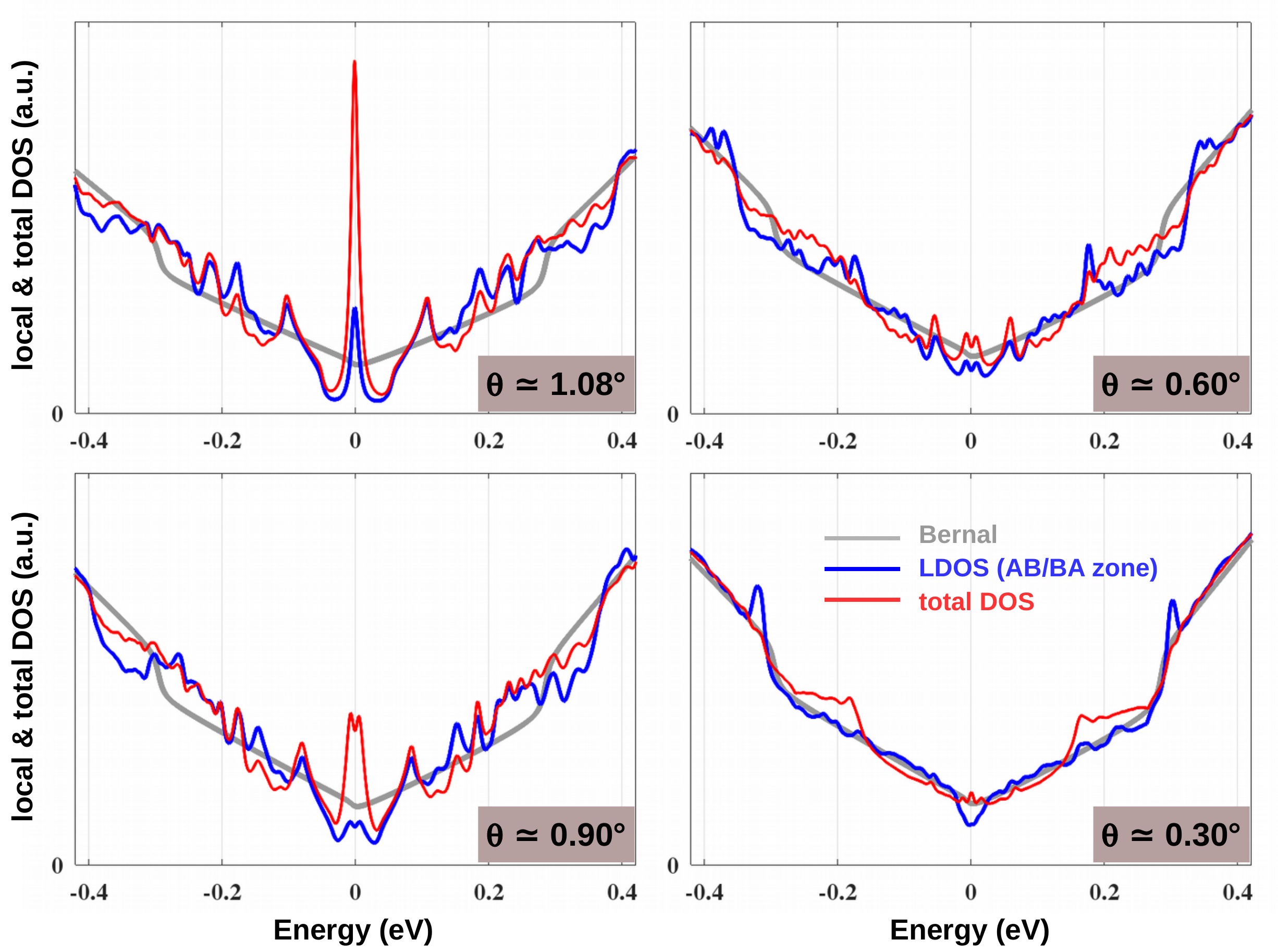}
	\caption{Local DOS in the AB/BA stacking domain and total DOS when decreasing $\theta$ in the small angle regime. The results obtained in Bernal stacking bilayer graphene are superimposed as a reference.}
	\label{fig_sim4}
	\end{figure}
	We now turn to the local electronic properties of TBLGs when the effect of the moir\'e superlattice relaxation is taken into account. The projected densities of states (PDOS) on carbon sites from different stacking domains and the corresponding local DOS (LDOS) are shown in Fig.~2. For large angles, only a weak spatial dependence is observed. However, this dependence is strongly enhanced for angles $\lesssim \theta^{\rm MA}_1$. More remarkably, in all small-angle cases shown in Fig.~2 (and more fully presented in Section 5 of the SI), the low-energy bands exhibit a strong localization in the AA stacking regions and, accordingly, a high LDOS peak near the Fermi level is consistently observed. Moreover, these localized bands are mostly isolated from high-energy bands by a few bands with relatively low LDOS in the energy range of $\sim [ \pm 0.1,\pm0.15]$ eV. Thus, the local electronic quantities in AA stacking domains of small-angle TBLGs represent all the typical characteristics of flat bands observed for the first magic angle. 
	These specific electronic properties due to the AA stacking regions do not occur in other domains.
		Specifically, the electronic structure of AB/BA stacking regions approaches that of Bernal stacking bilayer graphene when decreasing $\theta$ below $1.1^\circ$ (see Fig.~4) whereas helical edge states around SLs are observed around zero energy in TBLG systems with an extremely small twist angle (see Fig.~5). Note that these helical edge states are more pronounced in an enlarged (\textit{i.e.}, finite) range of energy only when a vertical electric field is applied (as shown in Ref. ~\onlinecite{huan18}, see additional discussions in Section 6 of the SI).
	In summary, small-angle TBLG systems present the following common features: (i) a strong spatial variation and (ii) a strong electron localization in AA stacking regions, similar to the associated flat bands observed for $\theta^{\rm MA}_1$.	
	\begin{figure*}[!t]
	\centering
	\includegraphics[width = 0.98\textwidth]{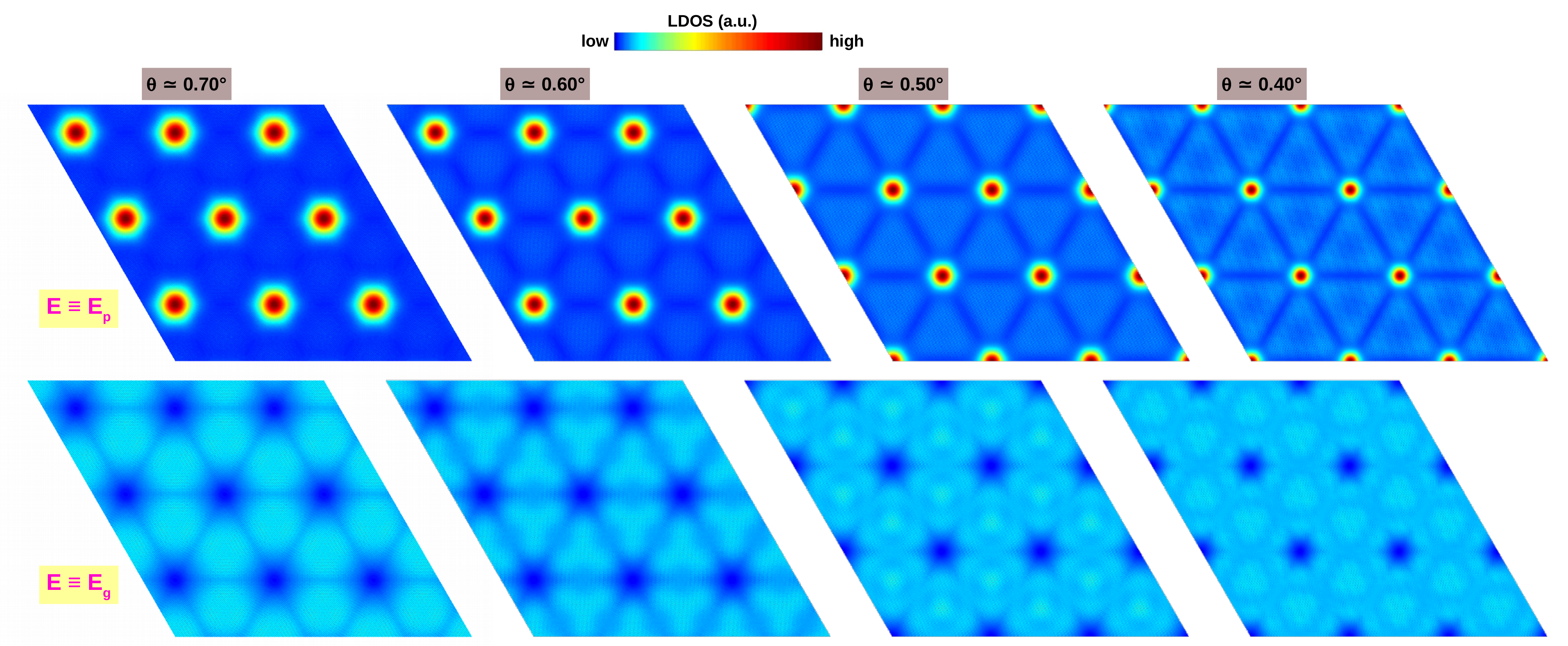}
	\caption{Local DOS at two energy points $E_p$ and $E_g$ marked in Fig.~2 are presented in the real-space for different twist angles.}
	\label{fig_sim5}
	\end{figure*}
	
	To further clarify the spatial dependence of the electronic properties of these small-angle TBLGs, the LDOS at the most strongly localized peaks in AA stacking regions (\textit{i.e.}, van Hove singularities at large angles and zero energy at small angles), are shown in their real space representation in Fig.~3(a). The inset is a typical real-space image depicting low-energy LDOS in these TBLG systems where a high peak is present in the AA regions and where the AB/BA zones and SLs are marked. Here, LDOS along a straight line from the center of an AA region to the center of the SL is investigated for different $\theta$ values. First, these results confirm that the spatial dependence and the electron localization in AA zones are weak for large twist angles but are strongly enhanced when reducing $\theta$. Most interestingly, the electron localization in AA stacking regions is found to saturate when $\theta$ passes below $\theta^{\rm MA}_1$. 
		This is further demonstrated by the real-space LDOS shown in Figs.~S3a and S3b as well as LDOS as a function of energy depicted in Fig.~2 and in Section 5 of the SI.
	Whereas other stacking domains with small LDOS at low energies are continuously enlarged when reducing $\theta$ in the regime $\lesssim \theta^{\rm MA}_1$, the electron localization peak in the AA stacking zones is almost unchanged. As a consequence, the contribution of AA stacking regions to the global properties of the TBLG system is monotonically reduced, thus explaining the disappearance of the localized peak in the total DOS at small twist angles as presented in Fig.~2 even though a highly localized peak in LDOS is still present in the AA stacking region. 
		The reduction of global electron localization when decreasing $\theta$ is demonstrated in more details in Fig.~4 and Figs.~S6a-S6b of the SI. Moreover, the global electronic structure of TBLG is shown to approach that of Bernal stacking bilayer graphene for very small angles. Small differences around $E \simeq \pm 0.2eV$ as seen for $\theta \simeq 0.3^\circ$ in Fig.~4 are due to the contribution of the electronic states around the soliton lines. This effect does not occur in the simple Bernal stacking system. These results imply that the electronic states from AB/BA stacking domains and around soliton lines present an increasingly dominant role to the global properties of the system when decreasing $\theta$ to small angles.
	Thus, we conclude here that although not observed directly in the global electronic quantities, the strong electron localization, similar to that associated to the presence of flat bands at $\theta^{\rm MA}_1$, is preserved locally in the AA stacking zones of TBLG systems below $\theta^{\rm MA}_1$.
	
		Alternatively, the absence of isolated flat bands (accordingly, strong global electron localization) in the small angle regime can be clarified by analyzing the bands connecting low-energy localized (really isolated and flat at $\sim 1.1^\circ$) bands with high energy ones. As shown and discussed in Fig.~2 and Figs.~S7a-S7b of the SI, they are the bands in the energy range of $\sim [ \pm 0.1,\pm0.15]$ eV. Real-space LDOS for two particular energy points marked in Fig.~2 is presented in Fig.~5 for different $\theta$ in the small angle regime. While the LDOS at $E_p$ confirms the presence of a strongly localized peak, the data at $E_g$ clearly exhibit a quasi electronic gap in the AA stacking zone. Most interestingly, the bands at $E_g$ (i.e., $\in [0.1,0.15]$ eV) present large LDOS values in the rest of the system, \textit{i.e.}, in the AB/BA domains and around soliton lines. This analysis confirms the increasing contribution of AB/BA stacking zones and soliton lines at small angles, thereby explaining why low-energy localized bands can not be as isolated and flat  as those observed at $\sim 1.1^\circ$.
	
	The variation (at large angles) and stability (at small angles) of the electron localization in AA zones presented in Fig.~3(a) are found to be essentially due to the dependence of stacking structure on the twist angle. This can be understood by analyzing the local structural deformations (\textit{i.e.}, the spatial variation of C-C bond length) for  different $\theta$ values (see Fig.~3(b)).
		Note that these local strain fields are a manifestation of both in-plane and out-of plane displacements (see Section~2 of the SI) and a direct consequence of the stacking structure of reconstructed TBLG superlattices as illustrated in the insets of Fig.~3(b).
	Our analysis shows the strong correlation between the electron localization in the AA stacking regions and the formation of those regions when varying the twist angle. In particular, in the large-angle ($> \theta^{\rm MA}_1$) regime, the AA stacking regions (see the images on top of Fig.~3(b)) and the electron localization (\textit{i.e.}, height and width of localization peak in Fig.~3(a)) are concurrently enlarged when reducing $\theta$. Both the size of the AA stacking regions and the electron localization peaks in these regions however reach a saturation point when approaching $\theta^{\rm MA}_1$. Indeed, they are almost unchanged (as illustrated in Fig.~3(a) and the images in the bottom of Fig.~3(b)) when further reducing $\theta$ in the small-angle ($\lesssim \theta^{\rm MA}_1$) regime. 
		All these features are illustrated further in the 2D real-space images of out-of-plane displacements and LDOS shown in Section 2 of the SI.
	The results presented in this section clearly establish the importance of atomic reconstruction in the emergence of the observed properties related to the electron localization in AA stacking regions as well as other common properties of small-angle TBLG moir\'e superlattice discussed above. 
	
	We will now discuss the consistency between the present theoretical results and the available experimental data from the literature. First, the first magic angle obtained here and accordingly the maximum electron localization at $\sim \theta^{\rm MA}_1$ are in very good agreement with the experimental measurements presented in Refs.~\cite{cao18a,kere19,gade20}, where the strongest many-body effects have been reported. Second, the real-space LDOS pictures as well as the structural properties presented 
		throughout the paper (most clearly, in Fig.~3 and Figs.~S3a-S3b of the SI)
	are consistent with Scanning Tunneling Microscopy and Spectroscopy (STM/STS) measurements~\cite{hude18,huan18,kere19,brih12,zhan20}. 	
	It is important to note that the size evolution (at large angles $> \theta^{\rm MA}_1$) and the stability (at small angles $\lesssim \theta^{\rm MA}_1$) of the AA stacking regions as well as of low-energy localization peak correspond to STM/STS results when varying the twist angle. More specifically, the width ($\sim 8 - 10$ nm) of the localization peak at small angles shown in Fig.~3(a) matches the experimental data presented in Refs.~\onlinecite{kere19,zhan20}. Finally, as emphasized above, one of the most important results is that even though it is not observed in the global electronic properties, the strong electron localization is preserved in the AA stacking regions of small-angle TBLG systems. This result is in agreement with the recent Raman study performed on small-angle TBLGs \cite{gade20}. In particular, this study shows that the full-width at half maximum of the G-band ($\Gamma_G$) measured by micro-Raman techniques is maximal around $\theta^{\rm MA}_1$ and  decreases monotonically when reducing $\theta$. These spectroscopic results thus confirm that the electron localization (the direct origin of large $\Gamma_G$) is globally maximal at $\theta^{\rm MA}_1$, but also demonstrate that such electron localization decreases monotonically for smaller values of $\theta$. 
		We note that the $\Gamma_G$ frequency is found to approach that in Bernal stacking bilayer graphene when decreasing $\theta$ to very small angles.
	Moreover, $\Gamma_G$ measured locally in the AA stacking regions of small-angle TBLGs (\textit{i.e.,} by the nano-Raman technique) remains large and similar to the $\Gamma_G$ value obtained at $\theta^{\rm MA}_1$, thus confirming the conservation of strong electron localization in the AA regions as presented.
		
	To summarize, the electronic properties of small-angle TBLGs with fully reconstructed moir\'e superlattices have been investigated using an atomistic tight-binding approach. The first magic angle obtained at $\theta^{\rm MA}_1 \simeq 1.1^\circ$ is found to be a critical case globally representing the maximum electron localization and, in effect, separates TBLGs into large- and small-twist angle classes with distinct electronic properties. In contrast to the large-angle cases where the model of low-energy Dirac fermions is conserved, small-angle TBLGs present strong spatial variation and strong electron localization (both in energy and real-space) in the AA stacking regions. In the small-angle regime, when $\theta$ is reduced below $\theta^{\rm MA}_1$, the relative contribution of AA stacking regions to the global properties of TBLG gradually reduces. Consequently, isolated flat bands as well as strong electron localization can no longer be observed in the global electronic properties of small-angle TBLGs, implying that magic angles below $1.1^\circ$ previously predicted by other theoretical models do not practically exist. These results are in very good agreement with existing experiments (both Raman and STM/STS spectroscopies) and thus provide a more comprehensive and accurate understanding of the electronic properties of small-angle TBLG systems. 
	
	\textbf{Acknowledgments} - V.-H.N. and J.-C.C. acknowledge financial support from the F\'ed\'eration Wallonie-Bruxelles through the ARC on 3D nano-architecturing of 2D crystals (N$^{\circ}$ 16/21-077), from the European Union’s Horizon 2020 Research Project and Innovation Program — Graphene Flagship Core3 (N$^{\circ}$ 881603), from the Flag-Era JTC project “TATTOOS” (N$^{\circ}$ R.8010.19), and from the Belgium FNRS through the research project (N$^{\circ}$  T.0051.18). V.M. acknowledges support from NY State Empire State Development’s Division of Science, Technology and Innovation (NYSTAR). Computational resources have been provided by the CISM supercomputing facilities of UCLouvain and the C\'ECI consortium funded by F.R.S.-FNRS of Belgium (N$^{\circ}$ 2.5020.11).

\bibliography{Nguyen_biblio}

\onecolumngrid
\includegraphics[page=1,scale=0.81]{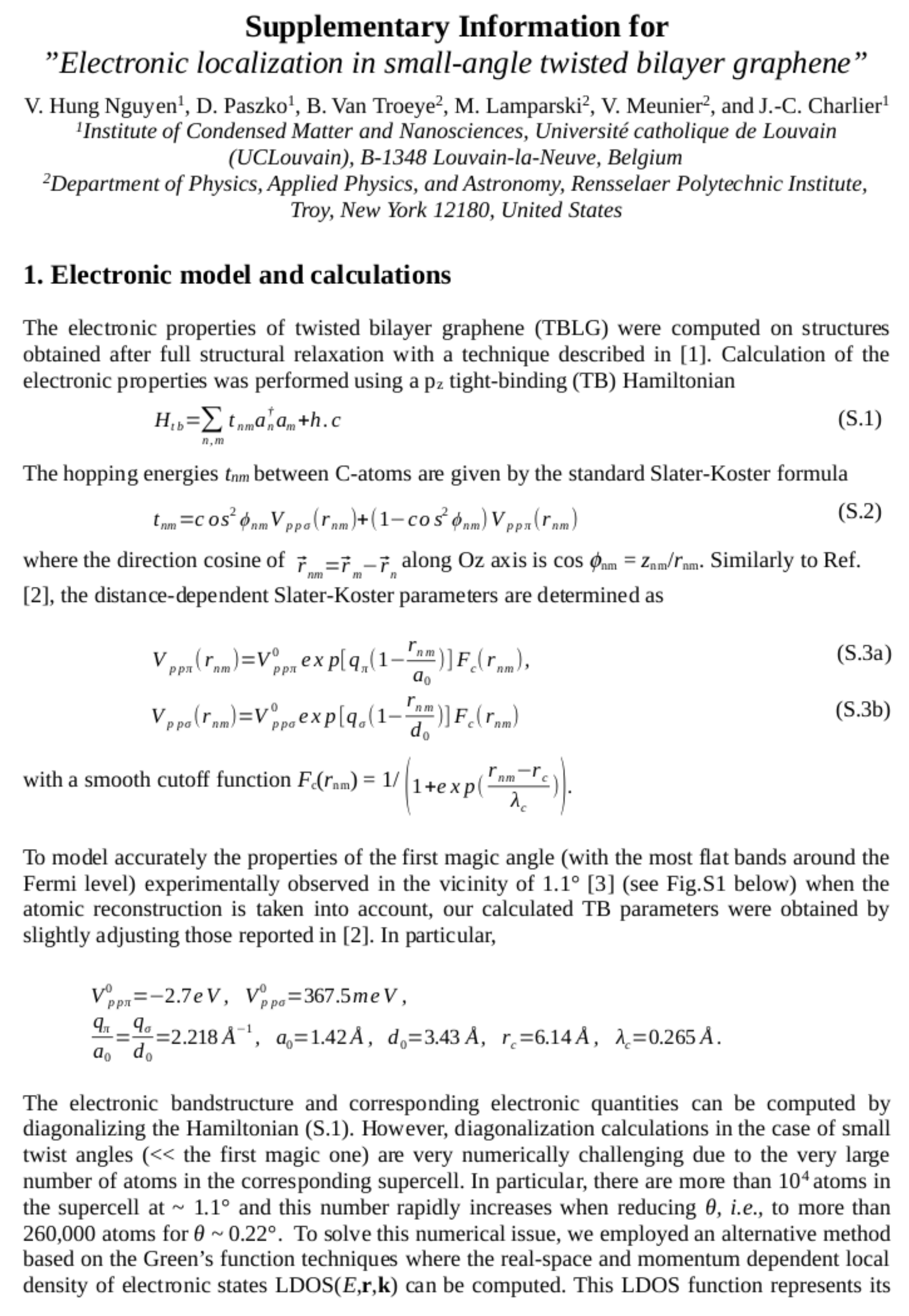} 
\includegraphics[page=2,scale=0.81]{SIarXiv.pdf} 
\includegraphics[page=3,scale=0.81]{SIarXiv.pdf} 
\includegraphics[page=4,scale=0.81]{SIarXiv.pdf} 
\includegraphics[page=5,scale=0.81]{SIarXiv.pdf} 
\includegraphics[page=6,scale=0.81]{SIarXiv.pdf} 
\includegraphics[page=7,scale=0.81]{SIarXiv.pdf} 
\includegraphics[page=8,scale=0.81]{SIarXiv.pdf} 
\includegraphics[page=9,scale=0.81]{SIarXiv.pdf} 
\includegraphics[page=10,scale=0.81]{SIarXiv.pdf} 
\includegraphics[page=11,scale=0.81]{SIarXiv.pdf} 
\includegraphics[page=12,scale=0.81]{SIarXiv.pdf} 
\includegraphics[page=13,scale=0.81]{SIarXiv.pdf} 
\includegraphics[page=14,scale=0.81]{SIarXiv.pdf} 

\end{document}